\documentstyle[epsfig,referee]{mn}{}
\begin{document}

\def\mpc{h^{-1} {\rm{Mpc}}}
\def\up{h^{-3} {\rm{Mpc^3}}}
\def\uk{h {\rm{Mpc^{-1}}}}
\def\lsim{\mathrel{\hbox{\rlap{\hbox{\lower4pt\hbox{$\sim$}}}\hbox{$<$}}}}
\def\gsim{\mathrel{\hbox{\rlap{\hbox{\lower4pt\hbox{$\sim$}}}\hbox{$>$}}}}
\def\kms {\rm{km~s^{-1}}}
\def\apj {ApJ}
\def\aj {AJ}
\def\aa {A \& A}
\def\mnras {MNRAS}

\title{Galaxy groups in the 2dF redshift survey: Galaxy Spectral Type
Segregation in Groups}

\author[M.J. Dom\'{\i}nguez et al]{
\parbox[t]{\textwidth}{
M.J. Dom\'{\i}nguez, A.A. Zandivarez, H.J. Mart\'{\i}nez, 
M.E. Merch\'an, H. Muriel \& D.G. Lambas}
\vspace*{6pt}\\ 
Grupo de Investigaciones en Astronom\'{\i}a Te\'orica y Experimental, 
IATE, Observatorio Astron\'omico, Laprida 854, C\'ordoba, Argentina \\
Consejo de Investigaciones Cient\'{\i}ficas y T\'ecnicas de la Rep\'ublica 
Argentina (CONICET)}
\date{\today}

\maketitle

\begin{abstract}
The behaviour of the relative fraction of galaxies 
 with different spectral types in groups is analysed
as a function of projected local galaxy density and the group-centric
distance.
The group sample was taken from the 2dF Group Galaxy Calatogue
constructed by Merch\'an \& Zandivarez.
Our group sample was constrained to have a homogeneous virial
mass distribution with redshift. 
Galaxies belonging to this group sample
were selected in order to minimize possible biases 
such as preferential selection of high luminosity objects.
We find a clear distinction between high virial
mass groups ($M_V\gsim 10^{13.5} M_{\odot}$) and the less massive ones.
While the massive groups show a significant dependence of the relative fraction
of low star formation galaxies on local galaxy density and group-centric radius,
groups with lower masses show no significant trends.
We also cross-correlate our group subsample with the previously identified 
clusters finding that this sample shows a very similar behaviour as observed
in the high virial mass group subsample.
 
\end{abstract}

\begin{keywords}
galaxies: groups - star formation - spatial distribution - segregation
\end{keywords}

\section{Introduction} 

There is a strong evidence that
high density environments can significantly affect many galaxy properties. 
This evidence is both theoretical (e.g. Gunn \& Gott 1972, Torman 1998)
and observational (e.g. Zabludoff \& Franx 1993, Henriksen \& Jones 1996).
In particular star formation 
rates (Dressler et al. 1985, Balogh et al. 1998, 
Allam et al. 1999, Hashimoto et al 1999, Moss \& Whittle 2000,
Carter et al. 2001), gas content (Giovanelli \& Haynes 1985,
Vollmer et al. 2001, Solanes et al. 2001) and
morphology change.
Dressler (1980) find a clear correlation between
galaxy morphology with the projected local galaxy density (hereafter 
$\Sigma_{gal}$) defined with the ten nearest galaxies on the sky.
Dressler et al. (1997) also found that, for distant clusters
the fraction of S0 galaxies is 2-3 times smaller than in lower redshift clusters
suggesting that S0's are generated in large numbers only after cluster
virialization.
Whitmore et al. (1993) reexamine Dressler's sample of galaxies in clusters
and suggest that the morphology-cluster centric distance relation is more
fundamental than the morphology-local galaxy density relation. 
 Whitmore \& Gilmore (1991) found that the fraction
of ellipticals is roughly $15\%$ at the edge of a cluster all the way
to about $0.5$ Mpc from the center, at which point it begins to rise 
dramatically, reaching values of $60-70\%$ at the very center.

Various physical process have been proposed to explain these and other 
systematic differences between the field and cluster galaxy populations.
Among them we can mention gas evaporation (Cowie \& Songaila 1977), 
ram pressure stripping (Gunn and Gott 1972, Abadi et al. 1999, 
Quilis et al. 2000), truncated star formation (Larson et al. 1980, Balogh,
Navarro \& Morris 2000), galaxy harassment (Moore et al. 1996), merging 
(Barnes 1992, Lavery \& Henry 1998, Steinmetz \& Navarro 2002), tidal striping 
and shaking (Bird \& Valtonen 1990, Miller 1988), etc. 
Although the above processes are all plausible, the effects provided by
initial conditions on galaxy formation could also be a very important part
in the segregation scenario within current hierarchical models of 
structure formation as the Cold Dark Matter model.

Recently, Dom\'{\i}nguez, Muriel \& Lambas  (2001) analysed the relative 
fraction of morphological galaxy types in clusters as a function
of the projected local galaxy density, and different cluster global
parameters: projected
gas density, projected total mass density and reduced clustercentric
distance. The authors conclude that there are different mechanisms
controlling the morphological segregation depending on the galaxy environment.
They found that mechanisms of global nature dominate in high density 
environments, namely the virialized regions of clusters, while local galaxy
density is the relevant parameter in the outskirts where the influence of
cluster as a whole is relatively small compared to local effects.

In the field, galaxies form stars at rates several times higher than systems
of similar luminosities at the cores of clusters. This is partly a result 
of the well-known morphology-density relation, since ellipticals and
S0 galaxies are more abundant in clusters (Dressler 1980), but there is evidence
that even later type galaxies in clusters and groups form stars at lower rates than
in the field (Zabludoff \& Mulchaey 1998, Balogh et al. 1999, Allam et al. 1999) 
suggesting that the cluster environment somehow curbs the star formation rates 
of all galaxies, regardless of morphology.

Postman \& Geller (1984) gave evidence for the existence of a morphology-density
relation in loose groups obtained from the CfA Redshift Survey. The authors
found an absence of morphology-density relation at very low densities.
However high density groups show a small change in the morphological fractions
for the densest bins. Whitmore (1995) suggests that a possible problem with
their study was the inclusion of clusters of galaxies in their sample.
Many of the galaxies in the densest groups are cluster members.
By removing these cluster galaxies the author reexamined the morphology-density
relation finding that it is very weak or non existent in groups.
Hashimoto et al (1999) using a very large and homogeneous dataset from
the Las Campanas Redshift Survey showed that the star formation rates
of galaxies are sensitive to the local galaxy density, in such
way that galaxies show higher levels of star formation in low density
than in high density environments. 

Most of the galaxies in the 
universe belong to groups of galaxies, but, due to the difficulty of
discerning them from the field, groups of galaxies are, as a whole, not as well
studied as larger systems. These favorable environments for galaxy 
interactions, where the influence of the intergalactic medium and the tidal
influences of the global potential are weaker, provide useful insights to 
understand the effects of the medium on morphology and star formation.
A useful parameter that  contains information on both morphology
and current star formation in galaxies is that defined in Madwick et al (2002)
where low values of this parameter, $\eta<-1.4$, correspond mainly 
to early-types dominated by an old stellar population while positive large
values of $\eta$ correlate with late morphological types and increasing 
star formation rates. 

Merch\'an \& Zandivarez (2002) have identified galaxy groups in the 2dF public 
100K data release using a modified Huchra \& Geller (1982) group finding algorithm
taking into account the 2dF magnitude limit and redshift completeness masks.
The global effects of group environment on star formation
was analysed by Mart\'{\i}nez et al (2002) using this catalogue.
They have found a strong correlation
between the relative fraction of different galaxy types and the parent group
virial mass. For groups with $M\gsim 10^{13} M_{\odot}$ the relative fraction
of star forming galaxies is significantly suppressed indicating that
the low mass group environment is affecting star formation.

In this paper, we present hints toward understanding local  
environment effects affecting the spectral types of galaxies
in groups, taking advantage of a very large and homogeneous available dataset.
In analogy with the dependence of morphology on environment
we introduce the spectral type-density relation and spectral type
fraction as a function of the group center distance for a subsample
of galaxies in groups taken from the catalogue of Merch\'an \& Zandivarez (2002).
The outline of this paper is as follows. Section 2 describes the selection
of the samples, whereas in section 3 the correlation between the fraction
of galaxies of different star forming characteristics with local
galaxy density and group center distance are analysed.
Finally, a summary is presented in section 4. 

\begin{figure}
\epsfxsize=0.5\textwidth
\hspace*{-0.5cm} \centerline{\epsffile{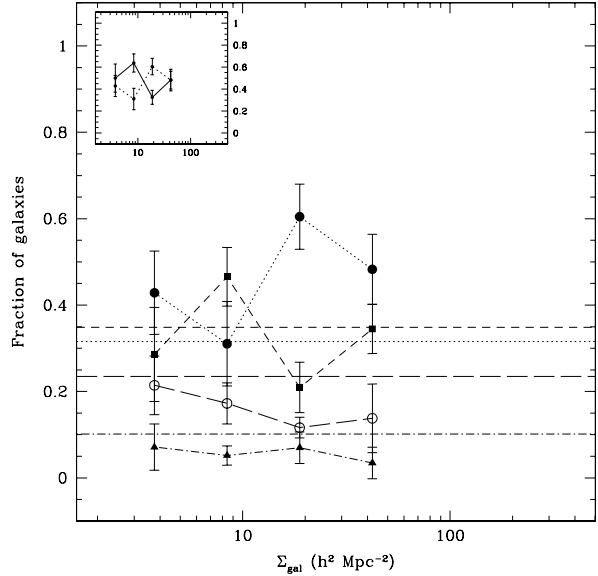}}
\caption{ 
Fraction of the different spectral types as a function of the projected local
galaxy density for galaxies in low mass groups ($M_V < 10^{13.5} M_{\odot}$). 
Type 1 correspond are plotted with filled circles connected with
dotted lines, Type 2 with filled squares connected with short dashed lines,
Type 3 with open circles connected with long dashed lines and Type 4 with
filled triangles connect with dotted dashed lines. Error bars were estimated 
using the bootstrap resampling technique. The horizontal lines are the mean
fraction of galaxies for each spectral type in the whole 2dFGRS with same 
selection criteria as galaxies in groups. In the small box in the upper left 
corner are displayed the fraction of spectral Type 1 (dotted line) and
Type 2 plus Type 3 (continuous line).
}
\label{fig1}
\end{figure}

\section{Sample selection}

Samples of galaxies and groups used in this work were selected from the  
Merch\'an \& Zandivarez (2002) group catalogue (hereafter 2dFGGC). 
This catalogue was constructed from the 2dF public 100K 
data release of galaxies with the best redshift estimates within the 
northern (NGP, $-37^{\circ}.5 \leq \delta \leq -22^{\circ}.5$, 
$ 21^h 40^m \leq \alpha \leq 3^h 30^m$) and southern 
($-7^{\circ}.5 \leq \delta \leq 2^{\circ}.5$; $9^h 50^m \leq \alpha 
\leq 14^h 50^m$) strips of the catalogue. 
The finder algorithm used for group identification is similar to that developed by Huchra \& Geller (1982) but modified taking into account 
redshift completeness and magnitude limit mask
present on the current release of galaxies
(see Figure 13 and 15 of Colless et al. 2001).
In the construction of the 2dFGGC values of $\delta \rho/\rho=80$ 
and $V_0=200~\kms$ were used to maximize the group accuracy. 
These optimal parameteres are the result of several tests using mock catalogues
that take into account the current radial and angular selection functions of
the 2dFGRS. The linking parameters were scaled to a fiducial redshift $cz=1000~ \kms$.
The resulting group catalogue comprises a total number
of 2209 galaxy groups with at least 4 members and
mean radial velocities in the range $900 ~\kms\leq V \leq 75000~\kms$.
Virial group masses were estimated using the virial radius and the
velocity dispersion ($M_{vir}=\sigma^2 R_V/G$, Limber \& Mathews 1960) where
the former is computed with the projected virial radius and the later 
with their radial counterpart. 
In Table 1 we show the comparison between the mean properties of
the 2dFGGC with the most recent results for groups in Las
 Campanas Redshift Survey (LCRS, Tucker et al 
2000) and the combination of the groups in the Updated Zwicky Catalogue and Southern Sky Redshift Survey 2 (UZC+SSRS2, Ramella et al 2002) 

\begin{table}
\tiny
\begin{center}
\caption{Comparison of Groups Catalogues Mean parameters.
UZC+SSRS2 values consider only groups with more than five members}
\begin {tabular}{ccccc}
\hline 
Catalogue & $N$ &  $\bar{\sigma}(km/s)$ & $\bar{M}(h^{-1} M_{\odot})$ & $\bar{R}_V(h^{-1}Mpc)$\\
\hline 
2dFGGC & $2209$ &  $261$ & $8.5\times 10^{13}$ & $1.12$\\
LCRSGC & $1495$ &  $164$ & $1.9\times 10^{13}$ & $1.16$\\
UZC+SSRS2 & $411$ &  $264$ & $4.6\times 10^{13}$ & $1.06$\\
\hline
\end{tabular}
\end{center}
\end{table}

Our sample of galaxies in groups is similar to the first sample used
in Mart\'{\i}nez et al. (2002), selected in order to
achieve the highest level of completeness.
The groups are limited to the redshift range $0.02\le z \le 0.056$ due to 
the highly homogeneous distribution of group virial masses with redshift. 
Therefore, we minimize the possibility of a preferential bias to high mass groups
in the sample. To prevent a sample biased to high luminosity galaxies,
we also introduce an absolute magnitude cut-off 
on the galaxies ($M_{b_J}\lsim -17.2$) defined by the volume sampled.
Thus any preferred galaxy type is avoided.

Madgwick et al (2002) have shown that for emission line galaxies,
the  equivalent width of $H_{\alpha}$
emission-line, $EW(H_{\alpha})$, is very tightly correlated
to the $\eta$ parameter defined in that work.
The $\eta$ parametrization of a galaxy spectral properties
is based upon a Principal Component Analysis of the galaxy spectra that takes
into account the relative emission/absorption line strength present 
in a galaxy's optical spectrum.
This classification correlates well with morphology and can be interpreted
as a measure of the relative current star-formation present in each galaxy.

Since we are interested in the study of the properties of galaxies
on systems of galaxies, we consider the 4 types of Madgwick et al (2002):
\begin{itemize}
\item Type 1: $~~~~~~~~~~\eta < -1.4$,
\item Type 2: $-1.4\leq \eta < ~~1.1$,
\item Type 3: $~~1.1 \leq \eta < ~~3.5$, 
\item Type 4: $~~~~~~~~~~\eta\ge ~~3.5$.
\end{itemize}
The Type 1 class is characterized with an old stellar population and
strong absorption features, the Types 2 and 3 comprise spiral
galaxies with increasing star formation, finally the Type 4
class is dominated by particularly active galaxies such as starbursts.
With this distinction we are able to analyse the environmental
dependence of galaxy spectral types in groups.

\begin{figure}
\epsfxsize=0.5\textwidth
\hspace*{-0.5cm} \centerline{\epsffile{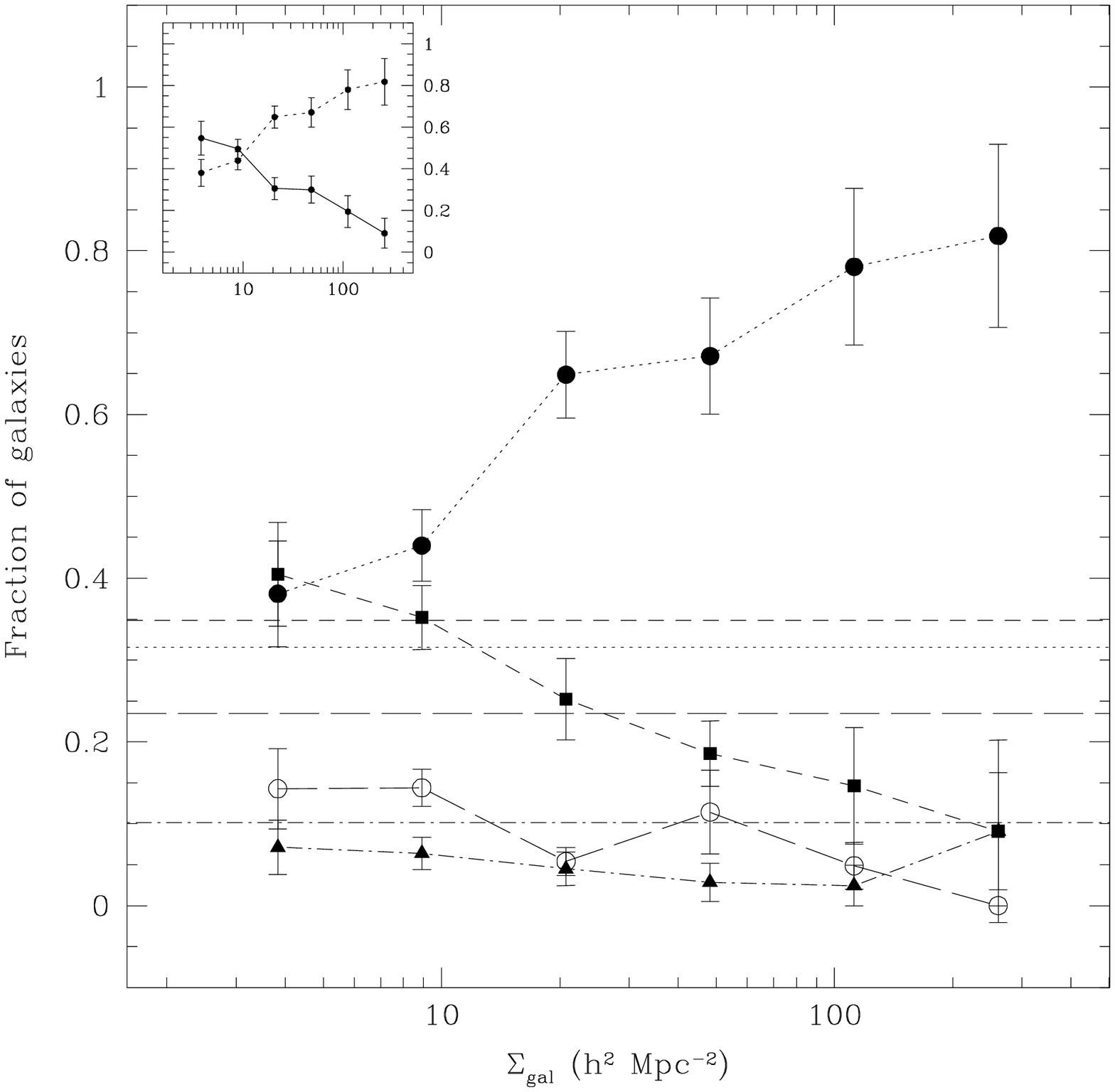}}
\caption{ 
Same as in Figure 1 but computed with galaxies in high mass groups
($M_V \ge 10^{13.5} M_{\odot}$).
}
\label{fig2}
\end{figure}

\begin{figure}
\epsfxsize=0.5\textwidth
\hspace*{-0.5cm} \centerline{\epsffile{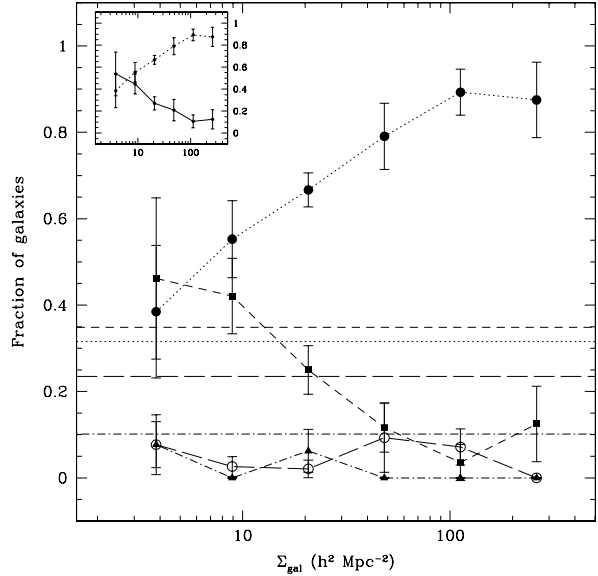}}
\caption{ 
Fraction of the different spectral types as a function of the projected local
galaxy density for galaxies in groups in the 2dFGGC previously identified as 
clusters (see Table 2). The symbols and lines are the same as in Figure 1.
}
\label{fig3}
\end{figure}

\section{Galaxy segregation analysis in groups}
In this section, the fraction of each galaxy type is studied
as a function of local galaxy environment, namely, the projected local galaxy
density and the normalized group-centric distance. 

\subsection{Projected local galaxy density}
We follow the suggestion of Dom\'{\i}nguez, Muriel \& Lambas (2001)
that at low density environments the local galaxy density is a primary
parameter in determining galaxy morphology. Therefore,
we analyse how different types correlate with projected local galaxy density.
This density was computed in a way similar to that in Dressler(1980)
 but using the area defined by the circle that encloses the fifth nearest 
neighbor to each galaxy. It is worth emphasizing the fact that since
each galaxy in 2dFGGC has redshift measurements so our statistical
analysis is free of projection effects.
We have restricted ourselves to those groups in our sample which have at least
8 members within the constraints described in
the previous section in order to improve the reliability of 
the statistical analysis. 
The stability of $\Sigma_{gal}$ was tested changing the number of neighbors
used in its computation. We have found no significant discrepancies in 
$\Sigma_{gal}$ using the fifth, sixth and seventh nearest neighbor.
The choice of the fifth member is due to the possibility of defining
a more reliable estimation of local density for the poorest groups that
dominate in number.
Another test which gives additional support to our choice of the fifth member
was performed in groups with more than 15 members computing $\Sigma_{gal}$
with the fifth and the tenth nearest neighbor as analysed  by Dressler (1980).
As in the previous test, we find that both computations of $\Sigma_{gal}$ are 
indistinguishable within the uncertainties. 

We split the sample into two subsamples of 
low ($M_V <10^{13.5} M_{\odot}$) and high 
($M_V \ge 10^{13.5} M_{\odot}$) virial mass. There are 18 and 32 groups
in each subsample with a total of 161 and 417 galaxies respectively.
In Figures 1 and 2 are shown the relative fraction of each galaxy type
as a function of $\Sigma_{gal}$ for low and high mass groups respectively.
Error bars in the figures were estimated using the bootstrap resampling technique. 
Horizontal lines in the figures are the mean fraction of galaxies for
different types within the 2dF Galaxy Redshift Survey
within the same sampled volume and luminosity cut-off.

By comparison of Figures 1 and 2 it can be appreciated that there is 
an important difference between the two group subsamples.
For the low-mass subsample there is no 
significant trend with $\Sigma_{gal}$. On the other hand, the high-mass
subsample exhibits a large increase in the fraction of  non star forming
galaxies (Type 1) with increasing $\Sigma_{gal}$.  
These behaviours can be better appreciated in the 
upper left panels in the Figures that correspond to the fractions of 
Types 1 (dotted line), and combined Types 2 and 3 (continuous line).
Type 4 galaxies, which correspond to the tail of the $\eta$ distribution
were omitted since they include particularly active galaxies and AGNs.   
These results are consistent with Whitmore (1995) and Maia \& da Costa (1990)
who also find a lack of a trend of morphology-density relation 
in groups when excluding clusters.

\begin{table*}
\begin{center}
\caption{Previously identified clusters in the subsample of the 2dFGGC}
\begin {tabular}{lccccc}
\hline 
\hline 
  Name & R.A.(degrees) & DEC (degrees) & z & $M_V$ ($h^{-1} M_{\odot}$) & $R_V$ 
($h^{-1} Mpc$) \\
\hline 
 APMCC 375    &  48.950 & -28.607 & 0.044 & 0.14E+14 & 0.47 \\
 ABELL 4049   & 357.133 & -28.460 & 0.029 & 0.71E+14 & 1.24 \\
 ABELL S1155  & 356.940 & -29.357 & 0.050 & 0.55E+14 & 0.77 \\
 ABELL S1171  & 359.728 & -27.723 & 0.028 & 0.58E+14 & 1.08 \\
 EDCC 155     & 337.220 & -25.629 & 0.034 & 0.16E+15 & 0.96 \\
 EDCC 129     & 334.070 & -24.640 & 0.038 & 0.16E+14 & 1.05 \\ 
 EDCC 121     & 333.441 & -25.430 & 0.031 & 0.22E+14 & 0.33 \\
 PCC N45-300  & 152.605 &  -2.410 & 0.043 & 0.34E+14 & 0.78 \\
 WBL 248      & 149.461 &  -2.736 & 0.020 & 0.36E+14 & 0.60 \\
 ABELL 0993   & 154.866 &  -4.691 & 0.055 & 0.24E+15 & 1.47 \\
 ABELL 0978   & 154.445 &  -6.133 & 0.055 & 0.26E+15 & 1.16 \\
 ABELL 1214   & 168.565 &  -5.253 & 0.039 & 0.24E+14 & 0.79 \\
 ABELL 1334   & 174.084 &  -3.993 & 0.056 & 0.20E+15 & 1.11 \\
 MKW 05       & 209.336 &  -2.760 & 0.025 & 0.44E+14 & 0.81 \\
 ABELL 0957   & 152.796 &  -0.688 & 0.045 & 0.36E+15 & 1.13 \\
 PCC N56-369  & 161.872 &   0.710 & 0.039 & 0.10E+15 & 1.13 \\
 ABELL 4053   & 357.983 & -27.840 & 0.050 & 0.26E+14 & 0.58 \\
 ABELL S0006  &   0.492 & -30.752 & 0.026 & 0.65E+13 & 0.48 \\
 ABELL S0001  & 359.973 & -30.853 & 0.030 & 0.54E+14 & 0.91 \\
 EDCC 694     &  41.476 & -27.991 & 0.023 & 0.19E+14 & 0.91 \\
\hline
\end{tabular}
\end{center}
\end{table*}

\begin{figure}
\epsfxsize=0.5\textwidth
\hspace*{-0.5cm} \centerline{\epsffile{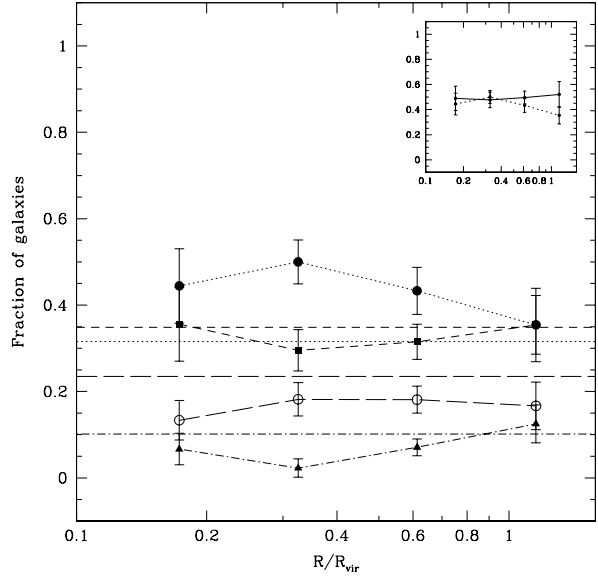}}
\caption{ 
Fraction of the different spectral types as a function of the groupcentric
distance normalized to the projected virial radius. The galaxies belong to
low mass groups ($M_V < 10^{13.5} M_{\odot}$).
The symbols and lines are the same as in Figure 1.
}
\label{fig4}
\end{figure}

\begin{figure}
\epsfxsize=0.5\textwidth
\hspace*{-0.5cm} \centerline{\epsffile{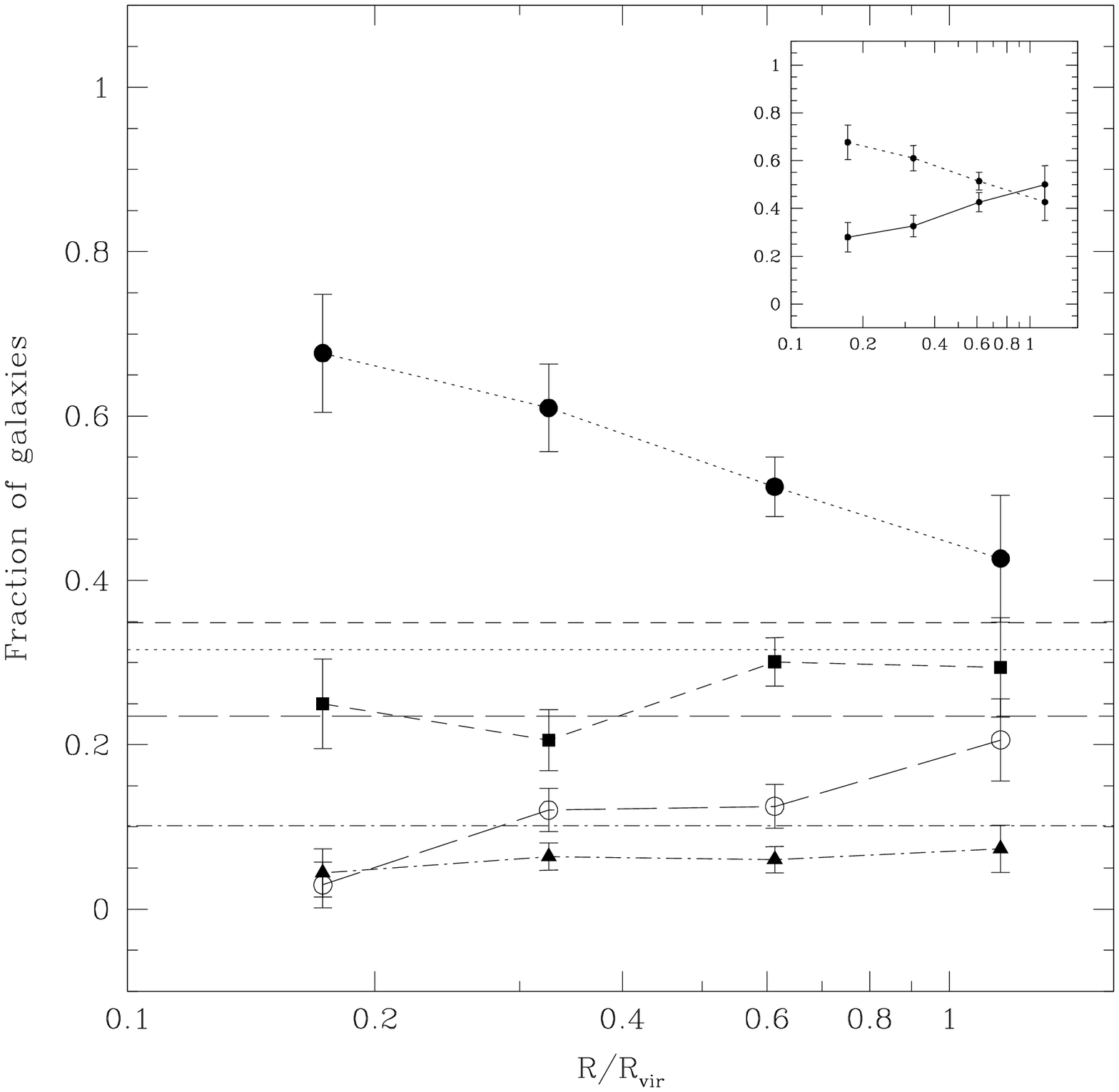}}
\caption{ 
Same as in Figure 4 but computed with galaxies in high mass groups
($M_V \gsim 10^{13.5} M_{\odot}$).
}
\label{fig5}
\end{figure}

\begin{figure}
\epsfxsize=0.5\textwidth
\hspace*{-0.5cm} \centerline{\epsffile{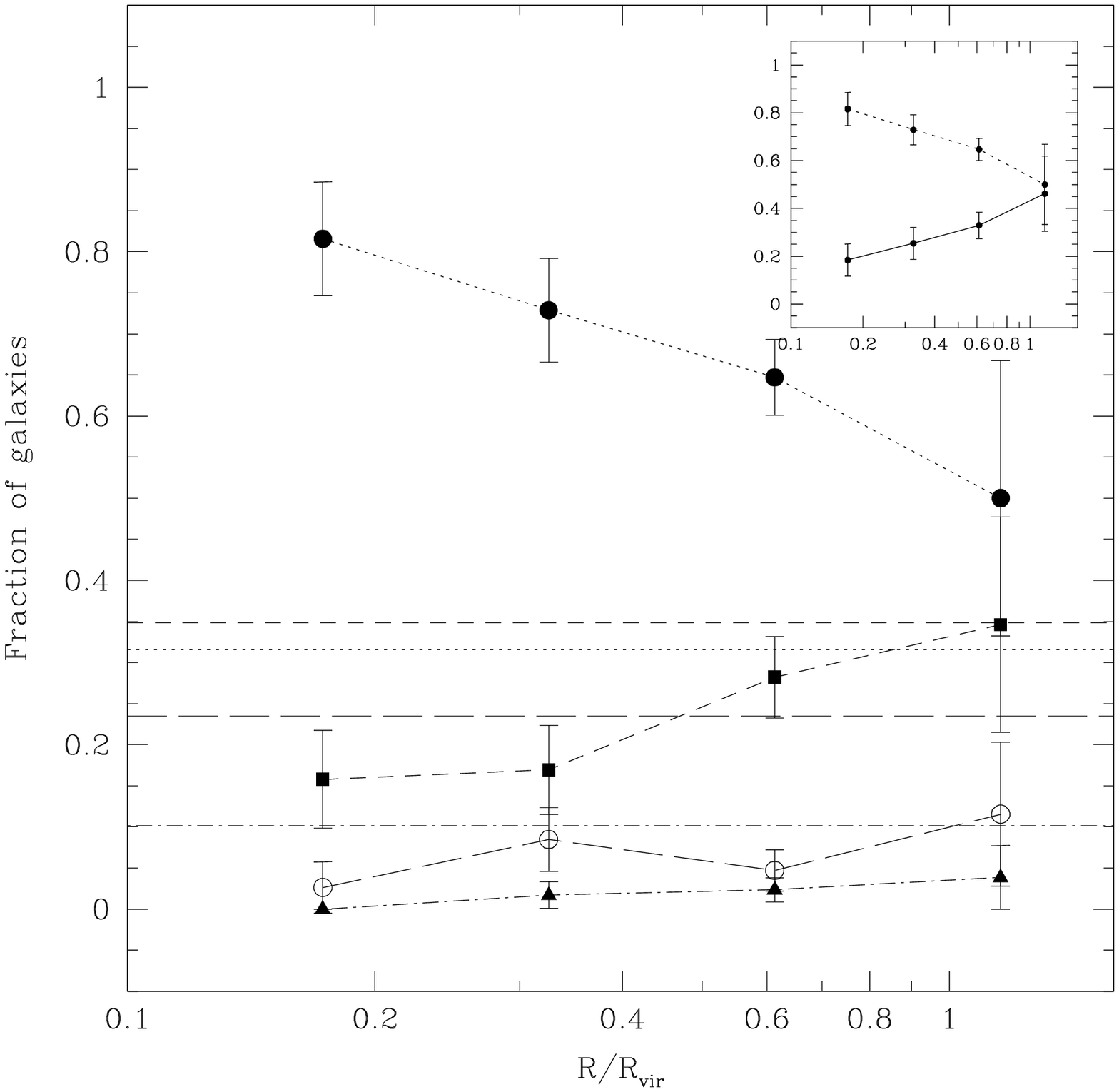}}
\caption{ 
Fraction of the different spectral types as a function of the groupcentric
distance normalized to the projected virial radius.
The galaxies belongs to groups in the 2dFGGC previously identified as 
clusters (see Table 2). The symbols and lines are the same as in Figure 1.
}
\label{fig6}
\end{figure}

We have also studied a sample consisting of groups which correspond to
clusters identified in previous surveys. By cross-correlating group
positions with clusters in the NASA/IPAC Extragalactic Database (NED), 
we have found 20 groups in this subsample of the
2dFGGC with a cluster identification in NED (see Table 2). 
When the same analysis is applied to this sample of groups (Figure 3)
we find a great similarity in the results to those found in the high
mass group subsample shown in Figure 2.
This is a totally consistent result since this sample is expected to
correspond to the most massive groups.

\subsection{Group-centric radius}
Another way to analyse the dependence of galaxy types within the group 
environments is through galaxy group-centric distances normalized to 
the group virial radius. In the analysis of Whitmore, Gilmore \& Jones (1993)
and Whitmore (1995) the cluster-centric radial distance is found to be a
 primary driver in contrast to the local galaxy density, a secondary parameter 
in the morphology segregation.
In this subsection we perform a similar analysis to that in subsection 3.1
using the group-centric distance instead of $\Sigma_{gal}$.
Since the computation of group-centric distances is less sensitive to the
number of galaxies than $\Sigma_{gal}$ we construct our sample with all
groups with at least 6 members. This choice allow us to improve the statistics.

We also split the sample into two subsamples of 
low ($M_V <10^{13.5} M_{\odot}$) and high 
($M_V \ge 10^{13.5} M_{\odot}$) virial mass as in subsection 3.1. 
There are 41 and 42 groups
in each subsample with a total of 308 and 494 galaxies, respectively.

In Figure 4 and 5 are shown the relative fraction of each galaxy type
as a function of the group-centric distance for low and high mass groups 
respectively.
Horizontal lines in the figures are the mean fraction of galaxies 
for the 2dF Galaxy Redshift Survey as in Figures 1 and 2. 

By comparison of Figures 4 and 5 it can be appreciated  
that there is a difference between the two group subsamples.
 For the low-mass subsample there is no significant trend 
with $R/R_{vir}$. On the other hand, the high-mass
subsample exhibits a continuous decrease of the fraction of non star forming
galaxies (Type 1) with increasing $R/R_{vir}$.  
These results are consistent with the results of Figures 1 and 2.
A possible explanation can be found in Balogh \& Navarro (2000) 
where it is found that star formation declines gradually after
galaxies enter in the system, as a result of the removal of 
the gaseous envelopes that supply the fuel needed for star formation.
This could explain the strong correlation found
by Mart\'{\i}nez et al (2002).  

An important analysis could be performed if X-ray information
were available for an important sample of the groups. Effects
associated with the intragroup medium might be  responsible of the removal of the
gas supply of the galaxies, in such X-ray group sample could provide
stronger gradient on galaxy fractions. 
The similarity of the gradients between the high mass subsample
and the previously identified clusters (showed in Figure 6) indicate 
the importance of a separate analysis for group and cluster environments.   

\section{CONCLUSIONS}
We have performed a correlation analysis between the relative fraction
of spectral types and the projected local galaxy density, $\Sigma_{gal}$,
in a sample of groups taken from the 2dFGGC constructed by Merch\'an \&
Zandivarez (2002).
Several possible sources of biases have been considered:
Firstly a reliable sample of groups has been selected
with a redshift independent virial mass distribution and 
a volume complete selection of their galaxy members.
Secondly, to test the stability of the results on $\Sigma_{gal}$
estimates, we have computed $\Sigma_{gal}$ with different
number of galaxy neighbours finding that
using the five nearest members gives accurate results. 
It should be remarked that our analysis was made on spectral types
of group member galaxies, so that this study in contrast to many
previous works lacks projection effects.
We find a clear distinction between high virial
mass groups ($M_V\gsim 10^{13.5} M_{\odot}$) and the less massive ones.
While the massive groups show a significant dependence of the relative fraction
of low star formation galaxies on $\Sigma_{gal}$,
groups with lower masses show no significant trends.

In a similar fashion, we have analysed the spectral type fractions as a function
of group-centric distance.
There is a significant difference between the behaviours
of the two subsamples.
While for the low-mass subsample there is no significant trend 
with $R/R_{vir}$, the high-mass
subsample shows a continuous decrease of the fraction of non star forming
galaxies (Type 1) with increasing $R/R_{vir}$.

In support of our analysis, we have considered  
a subsample of our groups that were previously identified 
as clusters. It is worth noticing that these objects are poor clusters
since they do not have strong X-ray emission. 
We find that this subsample 
is mainly composed of the tail of high group masses and
shows a very similar behaviour to our high virial mass samples.

\section*{Acknowledgments}
The authors thank the referee Douglas Tucker for helpful comments.
We thank the 2dFGRS for having made available the actual
data sets.
This research has made use of the NASA/IPAC Extragalactic Database 
(NED) which is operated by the Jet Propulsion Laboratory, California
Institute of Technology, under contract with the 
National Aeronautics and Space Administration. 
We thank E. Fitzgerald for suggestions.
This work has been partially supported by Consejo de Investigaciones 
Cient\'{\i}ficas y T\'ecnicas de la Rep\'ublica Argentina (CONICET), the
Secretar\'{\i}a de Ciencia y T\'ecnica de la Universidad Nacional de C\'ordoba
(SeCyT) and Fundaci\'on Antorchas, Argentina.

\end{document}